\documentclass[aps,pra,preprint,groupedaddress,showpacs]{revtex4}
\usepackage{amsmath,amssymb,amsthm,amsfonts}
\usepackage{graphicx}
\usepackage{bm}
\begin{document}
\title{Recent advances in description of few two-component fermions}

\author{O.~I.~Kartavtsev}
\author{A.~V.~Malykh}
\affiliation{Joint Institute for Nuclear Research, Dubna, 141980, Russia }

%\date{\today}

\begin{abstract}

Overview of the recent advances in description of the few two-component 
fermions is presented. 
The model of zero-range interaction is generally considered to discuss 
the principal aspects of the few-body dynamics. 
Particular attention is paid to detailed description of two identical 
fermions of mass $m$ and a distinct particle of mass $m_1$: it turns out that  
two $L^P = 1^-$ three-body bound states emerge if mass ratio $m/m_1$ increases 
up to the critical value $\mu_c \approx 13.607$, above which the Efimov effect
takes place. 
The topics considered include rigorous treatment of the few-fermion problem 
in the zero-range interaction limit, low-dimensional results, the four-body 
energy spectrum, crossover of the energy spectra for $m/m_1$ near $\mu_c $, 
and properties of potential-dependent states. 
At last, enlisted are the problems, whose solution is in due course. 

\end{abstract}

\maketitle

\section{Introduction}

\label{intro}

In the recent years, properties of multi-component ultra-cold quantum gases, 
including binary Fermi-Bose~\cite{Ospelkaus06,Karpiuk05,Fratini12} and 
Fermi~\cite{Shin06,Chevy06,Iskin06} mixtures and impurities 
embedded in a quantum gas~\cite{Cucchietti06,Kalas06,Mathy11} are under 
thorough experimental and theoretical investigation. 
In this respect, the low-energy few-body dynamics in two-species mixtures has 
attracted much attention. 
In particular, study of the energy spectrum and low-energy scattering 
of few two-component particles gives insight into the role of few-body 
processes in the many-body dynamics. 
One should also mention the reactions with negative atomic and 
molecular ions~\cite{Penkov99,Jensen03} and the two-component model for 
the three-body recombination near the Feshbach resonance~\cite{Kartavtsev02}. 

The aim of the paper is to present an overview of the recent advances in 
description of few ultra-cold two-component particles; it is assumed 
that identical particles are fermions interacting with a distinct particle 
via the $s$-wave potential, whereas interaction of identical fermions is 
forbidden. 
For the sake of generality, in some cases it is suitable to consider also 
a system of few non-interacting identical bosons and a distinct particle. 
It is worthwhile to mention that the $p$-wave interactions between fermions 
may be also important, in particular, an infinite number of the $1^+$ bound 
states was predicted~\cite{Macek06} for three identical fermions. 

To investigate the principal aspects of the few-body dynamics, it is natural 
to use the limit of zero-range two-body interaction, which allows one 
to obtain universal description of the few-body properties. 
In this respect, it is necessary to formulate rigorously the few-body problem 
in the zero-range interaction limit, which is an interesting problem in 
itself, especially if the identical particles are fermions. 

The zero-range model is suitably defined by imposing the boundary condition 
at the zero inter-particle distance. 
Thus, the interaction depends on a single parameter, e. g., the two-body 
scattering length $a$, which can be chosen as a length scale. 
As a result, the energy scale is $1/a^2$, and the only remaining parameter is 
mass ratio $m/m_1$, where $m$ and $m_1$ denote masses of identical and 
distinct particles, respectively. 
The units $\hbar = |a| = 2m/(1 + m/m_1) = 1$, for which the two-body binding 
energy $\varepsilon_2 = 1$, will be used throughout the paper. 

\section{Universal properties of three two-component fermions} 

\label{three-fermion} 

Investigation of two identical fermions and a distinct particle are of 
considerable interest for description of the multi-component ultra-cold gases. 
The three-body states of unit total angular momentum and negative parity 
($L^P = 1^-$) are especially important in treatment of the low-energy 
processes~\cite{Petrov03,Kartavtsev07}. 
A major progress was achieved in~\cite{Efimov73}, where it was shown that 
the zero-range model for sufficiently large mass ratio $m/m_1 > \mu_c$ does 
not provide unique description of the three two-component fermions. 
The unambiguous description of the three-body properties for $m/m_1 > \mu_c$ 
requires an additional parameter, which determines the wave function 
in the vicinity of the triple-collision point. 
As shown in~\cite{Efimov73}, a number of three-body bound states is infinite 
and their energies differ by the scaling factor. 
Later on, properties of the three-body spectrum were discussed also in 
a number of subsequent 
papers~\cite{Jensen03,Ovchinnikov79,Shermatov03,Li06,DIncao06}. 
The critical mass ratio is determined by a solution of the transcendental 
equation for $\sin\omega = 1/(1 + m_1/m)$ 
\begin{equation} 
\label{wc} 
\frac{\pi}{2} \sin^2\omega_c - \tan\omega_c + \omega_c = 0\, , 
\end{equation} 
which gives $\omega_c \approx 1.19862376 $ and 
$\mu_c = \sin\omega_c/(1-\sin\omega_c) \approx 13.6069657$. 
Furthermore, a significant achievement in this area was the construction of 
zero-energy solution for three two-component fermions in the interval 
$0 \le m/m_1 \le \mu_c$~\cite{Petrov03}, which provides the analytical 
expression for the low-energy recombination rate. 
A complete description of both the energy spectrum and the elastic and 
inelastic low-energy scattering is discussed below. 

\subsection{Hyperradial equations} 

In the zero-range limit, the two-body interaction is defined by imposing 
the boundary condition at the zero inter-particle distance $r$ 
\begin{eqnarray} 
\label{bound1} 
\lim_{r \rightarrow 0}\frac{\partial \ln (r\Psi)} {\partial r} = 
- \mathrm{sign} (a)\ . 
\end{eqnarray} 
The interaction introduced by means of the boundary condition is widely 
discussed in 
the literature~\cite{Demkov88,Wodkiewicz91,Idziaszek06,Kanjilal06}. 

Both qualitative and numerical results are obtained by using the solution 
of hyper-radial equations (HREs)~\cite{Macek68} 
\begin{equation} 
\label{system1}
%\fl
\left[\frac{d^2}{d \rho^2} - \frac{\gamma_n^2(\rho) - 1/4}{\rho^2} + E \right]
f_n(\rho) - \sum_{m = 1}^{\infty}\left[P_{mn}(\rho) - Q_{mn}(\rho)
\frac{d}{d\rho} - \frac{d}{d\rho}Q_{mn}(\rho) \right] f_m(\rho) = 0 \ ,
\end{equation}
whose terms $\gamma_n^2(\rho)$, $Q_{nm}(\rho)$, $P_{nm}(\rho)$ are derived 
analytically~\cite{Kartavtsev99,Kartavtsev06}. 
In fact, the critical mass ratio $\mu_c$ can be determined from the condition 
$\gamma_1(0) = 0$ as the first-channel effective potential at small $\rho$ 
takes the form $\left[ \gamma_1^2(0)- 1/4 \right] /\rho^2$, which implies that 
a number of the bound states is finite for $\gamma_1^2(0) > 0$ and infinite 
for $\gamma_1^2(0) < 0$. 

\subsection{Unit angular momentum}

\label{UnitL} 

\begin{figure}[hbt]
\includegraphics[width=.7\textwidth]{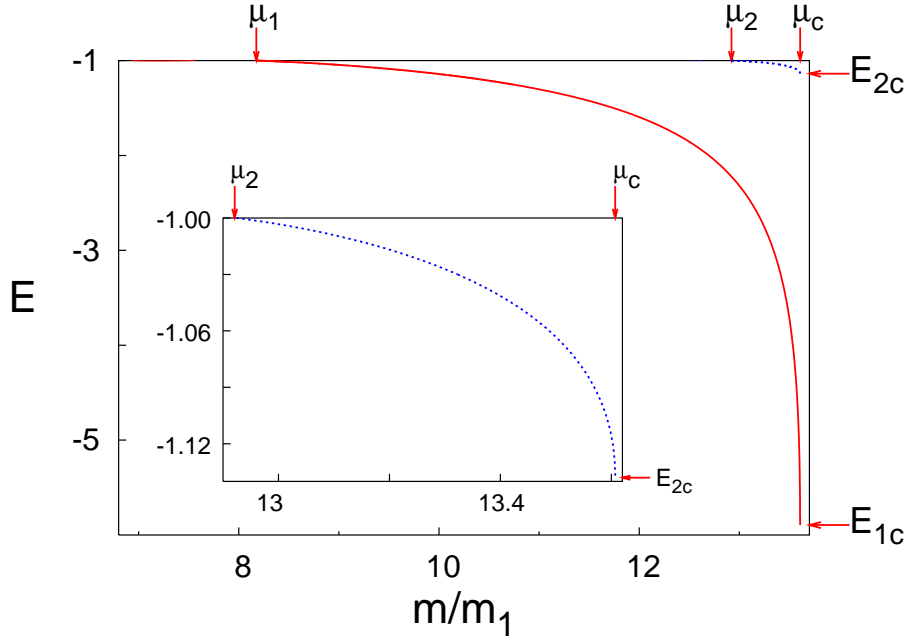} \\ 
{\caption{Dependencies of the bound-state energies (in units of the two-body
binding energy) on mass ratio $m/m_1$.
The arrows mark mass ratios $\mu_i$, for which the $i$th bound state
emerges from the two-body threshold, the critical mass ratio $\mu_c$, and
the bound-state energies $E_{ic}$ for $m/m_1 = \mu_c$.
In the inset the excited-state energy is shown on a large scale.} 
\label{fig}}
\end{figure}

Thorough studies of three two-component fermions in the states of total 
angular momentum and parity $L^{P} = 1^{-}$ were conducted 
in~\cite{Kartavtsev07}. 
Only negative parity is considered, since for the positive-parity states 
three particles do not interact by the s-wave zero-range potential. 
For the problem under consideration, the functions $\gamma_n(\rho)$ 
determining the effective potentials in~(\ref{system1}) satisfy 
the transcendental equation 
\begin{eqnarray} 
\label{transeq} 
\rho \, \mathrm{sign} (a) = \frac{1 - \gamma^2}{\gamma}\tan\gamma\frac{\pi}{2}
 - \frac{2}{\sin2\omega} \frac{\cos\gamma\omega}{\cos\gamma\frac{\pi}{2}} +
\frac{\sin\gamma\omega}{\gamma\sin^2\omega\cos\gamma\frac{\pi}{2}} \ .
\end{eqnarray}
In particular, taking the limit $\gamma_1(0)\to 0$ in Eq.~(\ref{transeq}), one 
obtains Eq.~(\ref{wc}) that determine $\mu_c$. 
From the solution of HREs~(\ref{system1}) it follows that for $a > 0$ there are no 
bound states in the interval $0 < m/m_1 < \mu_1$, exactly one bound state exists 
in the interval $\mu_1 \le m/m_1 < \mu_2$, and two bound states exist 
in the interval $\mu_2 \le m/m_1 \le \mu_c$, where $\mu_1 \approx 8.17260$ and 
$\mu_2 \approx 12.91743$. 
The bound-state energies decrease with increasing mass ratio on 
the interval $0 < m/m_1 \le \mu_c$, reaching the finite values 
$E_{1}(\mu_c) \approx -5.8954$, $E_{2}(\mu_c) \approx -1.13764$ at 
the critical value $m/m_1 = \mu_c $, and follow a square-root dependence 
$E_{i} - E_{i}(\mu_c) \propto \left(\mu_c - m/m_1 \right)^{1/2}$ near 
$m/m_1 = \mu_c $. 
The dependence of bound-state energies on mass ratio is illustrated in 
Fig.~\ref{fig}.

For mass ratio just below $\mu_i$ ($m/m_1 \lesssim \mu_1$ and 
$m/m_1 \lesssim \mu_2$), the relevant bound state turns to 
a narrow resonance, whose position $E^r_i$ continues a linear mass-ratio 
dependence of the bound-state energy, $E^r_i + 1 \propto \mu_i - m/m_1$, 
whereas the width $\Gamma_i$ depends quadratically, 
$\Gamma_i \propto (\mu_i - m/m_1)^2$. 

Calculations at the three-body threshold reveal the two-hump structure 
of the mass-ratio dependencies for the elastic (2 + 1)-scattering cross 
section and the three-body recombination rate. 
The dependence of the  recombination rate is in accordance with the analytical 
expression~\cite{Petrov03}. 
The structure of both isotopic dependencies stems from the interference of 
the incoming and outgoing waves in the 2 + 1 channel; the effect of 
interference is connected with emergence of two three-body bound states 
due to the deepening of the effective potential with increasing mass ratio. 
A detailed explanation of the mass-ratio dependence can be found 
in~\cite{Kartavtsev07}.

For negative scattering length $a < 0$ there are no three-body bound states 
in the interval $0 < m/m_1 \le \mu_c$. 

\subsection{Arbitrary angular momentum}

To obtain the comprehensive description of three two-component particles, it 
is necessary to study the three-body properties for arbitrary total angular 
momentum. 
In this respect, the three-body rotational-vibrational spectrum was analysed 
in~\cite{Kartavtsev07a}, where the identical particles were either fermions or 
bosons. 
It turns out that the properties of the three-body energy spectrum 
for arbitrary $L$ resemble those for $L = 1$~\cite{Kartavtsev07}. 

Method of calculations for arbitrary $L$ states in~\cite{Kartavtsev07a} was 
similar to that used for $L^{P} = 1^{-}$ states in~\cite{Kartavtsev07}, 
i.~e., the system of HREs whose terms $\gamma_n^2(\rho)$, $Q_{nm}(\rho)$, 
$P_{nm}(\rho)$ are derived analytically~\cite{Kartavtsev99,Kartavtsev06} was 
solved. 
By analysing the first-channel effective potential 
$(\gamma_1^2(\rho)-1/4)/\rho^2$, it was shown in~\cite{Kartavtsev07a} that 
the bound states exist if either the identical particles are fermions and $L$ 
is odd or the identical particles are bosons and $L$ is even. 
Thus, it is suitable to treat jointly both bosonic problem for even $L$ and
fermionic problem for odd $L$. 
The most discussed feature~\cite{Efimov73,Ovchinnikov79,Li06,DIncao06} of 
the three-body problem under consideration is an appearance of the infinite 
number of bound states for sufficiently large $m/m_1$. 
For each $L$, $\gamma_1^2(0)$ decreases with increasing $m/m_1$ and crosses 
zero at the critical value $\mu_{c}(L)$. 
Thus, the number of bound states is infinite for mass ratio above 
the critical value and finite below it. 
A set of critical mass ratios $\mu_{c}(L)$ are given in Table~\ref{tab3d} 
for $L = 1 - 5$. 
\begin{table}[htb]
\caption{Upper part: Mass ratios $\mu_N(L)$ at which the N-th bound 
state arises and the critical values $\mu_c(L)$. 
Lower part: Bound-state energies $|E_{L N} (\mu_c(L))|$ 
at critical mass ratios. }
\label{tab3d}
\begin{tabular}{lccccccc}
%$ L$  & \multicolumn{6}{c} {$\mu_N(L)$} & $\mu_c(L)$ \\
$ L$  & $N=1$ &  $N=2$ &  $N=3$ & $N=4$ & $N=5$ &
$N=6$ &  $\mu_c(L)$\\
\hline
 1 & 7.9300 & 12.789 & -      & -  & -  & -  & 13.6069657 \\
 2 & 22.342 & 31.285 & 37.657 & -  & -  & -  & 38.6301583 \\
 3 & 42.981 & 55.766 & 67.012 & 74.670 & -  & -  & 75.9944943 \\
 4 & 69.885 & 86.420 & 101.92 & 115.08 & 123.94 & -  & 125.764635 \\
 5 & 103.06 & 123.31 & 142.82 & 160.64 & 175.48 & 185.51 & 187.958355 \\
\hline
 1 & 5.906 & 1.147 & -     & -     &    -  &-&\\
 2 & 12.68 & 1.850 & 1.076 & -     &    -  &-&\\
 3 & 22.59 & 2.942 & 1.417 & 1.057 &    -  &-&\\
 4 & 35.59 & 4.392 & 1.920 & 1.273 & 1.049 &-&\\
 5 & 52.16 & 6.216 & 2.566 & 1.584 & 1.206 & 1.045 & \\
\hline
\end{tabular}
\end{table}

It was shown that three particles are not bound for $a < 0$ and mass ratio 
in the interval $m/m_1 \le \mu_{c}(L)$. 
There is a finite number of bound states for $a > 0$ and 
$m/m_1 \le \mu_{c}(L)$. 
More precisely, three particles are unbound for sufficiently small $m/m_1$ 
and with increasing mass ratio the $N$th bound state arises 
at $m/m_1 = \mu_N(L)$. 
The energies of each bound state $E_{L N}(m/m_1)$ monotonically decrease with 
increasing mass ratio, reaching the finite values at $\mu_c(L)$ and 
obeying the $\left[\mu_c(L) - m/m_1 \right]^{1/2}$ dependence just below 
$\mu_{c}(L)$. 
For $m/m_1$ just below the critical values $\mu_N(L)$, there are resonances, 
whose positions depend linearly and widths depend quadratically on 
the mass-ratio excess $\mu_N(L) - m/m_1$. 

Reasonable estimates (up to a few percent accuracy) of bound-state energies 
for $a > 0$ were obtained numerically by using the one-channel approximation 
for the total wave function. 
Critical values of mass ratios $\mu_N(L)$ and three-body energies 
$E_{L N}(m/m_1)$ at $m/m_1 = \mu_{c}(L)$ are  given in Table~\ref{tab3d}.
Accuracy of the one-channel approximation can be estimated by comparing 
the $L = 1$ values $\mu_N(1)$ and $E_{1 N}[\mu_{c}(1)]$ from Table~\ref{tab3d} 
with the precise values $\mu_1$, $\mu_2$, $E_{1}(\mu_c)$, and $E_{2}(\mu_c)$ 
given in Section~\ref{UnitL}. 
Notice that for $L = 3-5$ the uppermost bound states are loosely bound and 
appear very close to the corresponding critical values $\mu_{c}(L)$. 
Thus, taking into account an estimated accuracy of the approximation, one 
concludes that more careful calculation is necessary to describe these loosely 
bound states. 

As it was shown in~\cite{Kartavtsev07} for $L = 1$, an appearance 
of the three-body bound states with increasing mass ratio is intrinsically 
connected with the oscillating behaviour of the $2 + 1$ elastic-scattering 
cross section and the three-body recombination rate. 
Analogously, the dependence of the scattering amplitudes on mass ratio for 
higher $L$ would exhibit the number of interference maxima which are related 
to the appearance of up to $N_{max}$ bound states. 

A reasonably good description of the energy spectrum is obtained within 
the framework of the quasi-classical approximation, which allows 
the asymptotic expression of bound-state energies $E_{L N}(m/m_1)$ to be 
derived for large $L$ and $m/m_1$. 
As a result for $\mu_{1}(L) \le m/m_1 \le \mu_{c}(L)$ and $a > 0$, all 
the bound-state energies are described by the universal function of two scaled 
variables $\xi = (N - 1/2)/\sqrt{L(L + 1)}$ and 
$\eta = \sqrt{m/[m_1 L (L + 1)]}$. 
This scaling dependence is confirmed by the numerical calculations for $L > 2$ 
and is in good agreement even for small $L = 1, 2$. 
The universal description implies that $\mu_c(L) \approx 6.218(L + 1/2)^2$ 
and $\mu_1(L) \approx 3.152 (L + 1/2)^2$ for large $L$, while 
the number of vibrational states for given $L$ is limited as 
$N_{max} \le 1.1 \sqrt{L(L + 1)} + 1/2$. 
More details are presented in paper~\cite{Kartavtsev07a}. 

\subsection{Integral equations} 

Description of the two-component systems was obtained also by solving 
the momentum-space integral equations (generalised Skorniakov-Ter-Martirosian 
equations) in~\cite{Endo11,Helfrich11}. 
These calculations generally confirm the results 
of~\cite{Kartavtsev07,Kartavtsev07a} and contain some additional details. 
One should mention that the values $\mu_N(1)$ calculated in~\cite{Endo11} 
are in agreement with those obtained in~\cite{Kartavtsev07} for $L = 1$, 
while $\mu_N(L)$ from~\cite{Endo11} for $L = 2 - 4$ slightly differ from 
the approximate values given in~\cite{Kartavtsev07a}. 
Besides, the characteristics of ($2 + 1$) zero-energy scattering for 
$L = 0 - 3$ were obtained in~\cite{Endo11}. 
These calculations provide an additional evidence for the appearance 
of three-body bound states, in particular, the $P$-wave scattering volume 
diverges exactly at $m/m_1$ tending to the critical values $\mu_1$ and 
$\mu_2$. 
The mass-ratio dependence of the elastic $(2 + 1)$-scattering cross sections 
for different energies below the three-body threshold was studied 
in~\cite{Helfrich11}. 
In addition, the momentum-space integral equations were applied to solution of 
different few-body 
problems~\cite{Levinsen09,Levinsen11,Combescot12,Alzetto12,Endo12}. 
More details of these calculations are given in Section~\ref{conclusion}. 

\subsection{Two-dimensional problem}

Study of three two-component particles with zero-range interactions confined 
in two dimensions was performed in~\cite{Pricoupenko10}, where the mass-ratio 
dependence of the three-body energies and a set of critical values $m/m_1$, 
at which bound states emerge, were obtained by solving the momentum-space 
integral equations. 
Similar to 3D problem~\cite{Kartavtsev07a}, the bound states exist if either 
the identical particles are fermions and $L$ is odd or the identical particles 
are bosons and $L$ is even; the binding energies monotonically increase with 
increasing mass ratio. 
Calculation~\cite{Pricoupenko10} shows that for $L = 0$ two bosons and 
the third particle are bound for any mass ratio, while the second and third 
bound state appear at $m/m_1 \approx 1.77$ and $m/m_1 \approx 8.341$. 
Likewise, for $L = 1$ two fermions and the third particle are bound for 
$m/m_1 \ge  3.34$, while the second and third bound state appear 
at $m/m_1 \approx 10.41$ and $m/m_1 \approx 20.85$. 
As in 3D problem~\cite{Kartavtsev07a}, different rotational-vibrational states 
become quasi-degenerate for large $L$ and $m/m_1$. 

More general problem for three two-component particles was considered in 
paper~\cite{Bellotti11}, where all three zero-range interactions were taken 
into account. 
In particular case of two noninteracting identical particles, a set 
of mass-ratio values, at which the $L = 0$ three-body bound states emerge, are 
consistent with those found in~\cite{Pricoupenko10}. 
Furthermore, the energy spectrum of three 2D particles for different 
combinations of all masses and interaction strengths was considered 
in~\cite{Bellotti12}. 
The $S$- and $P$-wave elastic $(2 + 1)$-scattering and the three-body 
recombination in 2D two-component mixtures were studied for mass ratio 
corresponding to both $^6\mathrm{Li}$--$^{40}\mathrm{K}$ and equal-mass 
particles~\cite{Ngampruetikorn12}. 
A number of 2D bound-state and scattering properties for three and four 
particles were calculated in~\cite{Brodsky06}, where different combinations 
of equal-mass bosons and fermions were considered. 

The transition from two dimensions to three dimensions was studied in 
a quasi-two-dimensional geometry by confining particles in a harmonic 
potential along one direction~\cite{Levinsen12}. 
It was shown that $P$-wave energies of two identical fermions and one 
distinct particle smoothly evolve from $3D$ to $2D$ (with increasing confining 
frequency) for $m/m_1 \le \mu_c$. 
Correspondingly, the mass ratio, at which the three-body bound states emerges, 
increases from $2D$ value $3.33$~\cite{Pricoupenko10} to $3D$ 
value $8.17260$~\cite{Kartavtsev07}. 
In addition, it was estimated in~\cite{Levinsen12} that in $2D$ limit three 
identical fermions and one distinct particle are bound for $m/m_1 > 5$. 

\subsection{One-dimensional problem}

Properties of three two-component particles confined in one dimension with 
contact ($\delta$-function) interactions were considered 
in~\cite{Kartavtsev09}. 
It is assumed that attractive interaction of strength $\lambda < 0$ acts 
between each of two identical particles (either bosons or fermions) and a 
distinct one, while the strength of interaction between the identical 
particles is arbitrary $\lambda_1$. 

The three-body energy spectrum and the scattering length $A$ for collision 
of a bound pair off the third particle were calculated for two values 
$\lambda_1 = 0$ and $\lambda_1 \to \infty$ of the even-parity bosonic problem. 
Two sets of mass-ratio values, at which the three-body bound states arise and 
at which $A = 0$, were calculated both for $\lambda_1 = 0$ and 
$\lambda_1 \to \infty$. 
It is important to recall that the bosonic problem for $\lambda_1 \to \infty$ 
is exactly equivalent to the fermionic problem, for which the contact 
interaction between fermions is absent ($\lambda_1 = 0$). 
For the odd-parity states it was shown that three particles are not bound and 
the $(2 + 1)$ - scattering length for $\lambda_1 = 0$ was calculated. 

In addition, few analytical results were presented, in particular, it was 
shown that exactly one bound state of three equal-mass particles ($m/m_1 = 1$) 
exists for arbitrary $\lambda_1$. 
Next, for two light fermions (in the limit of $m/m_1 \to 0$) three particles 
are not bound for sufficiently large repulsive interaction between fermions, 
viz., for $\lambda_1/|\lambda|$ above the critical value 
$\approx 2.66735$~\cite{Cornean06}, and exactly one bound state exists for 
$\lambda_1/|\lambda|$ below this value. 
To elucidate the general features of three one-dimensional particles, both 
numerical and analytical results were used to construct a schematic ``phase'' 
diagram, which shows the number of three-body bound states and a sign of 
the $(2 + 1)$-scattering length $A$ in the plane of the parameters $m/m_1$ and 
$\lambda_1/|\lambda|$. 

\subsection{Dimensional analysis}

Contrary to the three-body problem in 3D, the 2D and 1D solutions remain 
regular near the triple-collision point even in the limit of zero-range 
two-body interaction; therefore, there is neither Thomas nor Efimov effect. 
As a result, it is not necessary to introduce an additional regularisation 
parameter and the low-energy three-body properties in 2D and 1D are completely 
determined by the two-body input. 

The results of
calculations~\cite{Kartavtsev09,Pricoupenko10,Kartavtsev07,Kartavtsev07a} 
give an opportunity to analyse the dependence of the three-body low-energy 
properties on the configuration-space dimension. 
It turns out that two identical fermions and a distinct particle are bound 
in $1D$ for $m/m_1 \ge 1$, in $2D$ for $m/m_1 \ge 3.33$ and in $3D$ for 
$m/m_1 \ge 8.17260$. 
Two identical non-interacting bosons and a distinct particle are bound in $1D$ 
and $2D$ for any mass ratio, the first excited state appears in $1D$ 
at $m/m_1 \approx 2.869539$ and in $2D$ at $m/m_1 \approx 1.77$, while in $3D$ 
the number of bound states is infinite. 
If the three-body binding energy exceeds the two-body binding energy, 
the production of the triatomic molecules becomes energetically more 
favourable than diatomic ones. 
For two identical fermions and a distinct particle it is justified if 
$m/m_1 > 49.8335$ in $1D$, $m/m_1 > 18.3$ in $2D$, and $m/m_1 > 12.69471$ in 
$3D$. 

\section{Potential-dependent states} 

Analysing the few-body properties for the small interaction range $r_0$ 
tending to zero, it is necessary to take into account two different length 
scales, $r_0$ and $a$ ($r_0<<a$), which means that all the states should be 
classified as either universal ones, whose energies scale according to 
$a^{-2}$ or potential-dependent ones, whose energies scale according to 
$r_0^{-2}$. 
In the unitary limit $a \to \infty$, the energy of universal state tends to 
zero, whereas the energy of potential-dependent state remains finite. 

The three-body bound states for two-component fermions in the limit of 
the infinite two-body scattering length were considered 
in~\cite{Blume10,Blume10a}. 
The interaction between different particles was taken as the Gaussian 
potential, whose parameters were adjusted to provide $a \to \infty$. 
It was found that in the zero interaction-range limit the $L^{P} = 1^{-}$ 
three-body bound state arises for mass ratio above $\approx 12.314$. 
This value is close to $m/m_1 \approx 12.31310$ determined from the condition 
$\gamma_1(0) = 1/2$, which means that above this mass ratio the first-channel 
effective potential in~(\ref{system1}) 
$\left( \gamma_1^2 - 1/4 \right) /\rho^2$ becomes attractive at small $\rho$. 
A similar problem was considered in~\cite{Gandolfi10}, in which 
the potential-dependent three-body bound states were found to exist at least 
for $m/m_1 > 13$ and two forms of the two-body potential. 

\section{Crossover at the critical mass ratio}

As discussed in Section~\ref{three-fermion}, in the zero-range limit 
the three-body properties are essentially different for mass 
ratio below and above the critical value $\mu_c$, e.~g., a number of bound 
states abruptly increases from two to 
infinity~\cite{Kartavtsev07,Kartavtsev07a,Endo11,Helfrich11}. 
Thus, one naturally needs to describe a crossover with increasing $m/m_1$ 
beyond the critical value $\mu_c$. 

Recently, to study the crossover at $\mu_c$, the dependence of the three-body 
bound state energies on mass ratio and the additional short-range three-body 
potential was calculated by using the momentum-space integral equations 
in~\cite{Endo12}. 
For mass ratio above $\mu_c$ the three-body potential is necessary to provide 
unambiguous description of the wave function near the triple collision point. 
Explicitly, the momentum cutoff in the integral equation is introduced 
in paper~\cite{Endo12} by using the dimensionless parameter $\Lambda$. 
Analysing the dependence of three-body energies in the plane of two parameters 
$m/m_1$ and $\Lambda$, it was found that the universal bound states found 
in~\cite{Kartavtsev07,Kartavtsev07a,Endo11,Helfrich11} are located in the 
region $\mu_1 \le m/m_1 \le \mu_c$ and $\Lambda^{-1} \to 0$, whereas 
the Efimov states are located  in the region $m/m_1 > \mu_c$ and 
$\Lambda^{-1} \to 0$. 
These two regions are joined by the crossover area, where the bound-state 
energies depend on both $m/m_1$ and $\Lambda$ and are almost independent 
of the particular choice of the three-body potential. 

\section{Zero-range model in the few-fermion problem}

\label{zero-range-model} 

Mathematical aspects of application of the zero-range-interaction model 
to few two-component fermions were investigated, e. g., 
in papers~\cite{Shermatov03,Minlos11,Minlos12,Correggi12}. 
In this respect, note the paper~\cite{Minlos11}, in which it was shown that  
the zero-range model for three and four fermions can be correctly defined 
for sufficiently small mass ratios. 
Similar statement was made for an arbitrary number of fermions 
in~\cite{Correggi12}. 
Among discussions of properties of three two-component fermions, one should 
mention a paper~\cite{Shermatov03}, in which the original Efimov's statement 
was proved, viz., the energy spectrum for the quantum numbers $L^{P} = 1^{-}$ 
becomes unbound from below for $m/m_1 > \mu_c$ (corresponding to 
$\gamma_1^2(0) < 0$). 
Recently, it was shown~\cite{Minlos12} that the three-fermion Hamiltonian is 
ambiguous in the zero-range limit if mass ratio exceeds $12.31310$ 
(corresponding to $\gamma_1(0) \le 1/2$). 
It is important to note simple considerations~\cite{Nishida08}, 
which indicates an existence of two square-integrable solutions in 
the vicinity of the triple-collision point $\rho \to 0$ for mass ratio 
in the interval $8.619 \le m/m_1 \le 13.607$ (corresponding to 
$1 \ge \gamma_1(0) \ge 0$). 
Therefore, for mass ratio that belongs to this interval, special care is 
needed to describe three two-component fermions. 

\section{Concluding remarks}

\label{conclusion}

An appearance of three-atomic molecules containing two heavy and one light 
particles crucially determines the equilibrium states and dynamics of both 
fermionic and fermionic-bosonic mixtures. 
One of interesting examples is the mixture of strontium and lithium 
isotopes. 
As for the $^7\mathrm{Li}$ -- $^{87}\mathrm{Sr}$ mixture mass ratio 
($m/m_1 \approx 12.4$) gets between $\mu_1$ and $\mu_2$ (at which the first 
and second bound states emerge), one expects that there is exactly one 
$P$-wave bound state of $^7\mathrm{Li} \, ^{87}\mathrm{Sr}_2$ molecule. 
Calculations~\cite{Kartavtsev07} predict energy of this molecule about
$-1.793$ (recall that the binding energy of the diatomic $^7\mathrm{Li} \,
 ^{87}\mathrm{Sr}$ molecule is taken as energy unit). 
For the $^6\mathrm{Li}$--$^{87}\mathrm{Sr}$ mixture mass ratio 
($m/m_1 \approx 14.5$) slightly exceeds $\mu_c$, which means that there are 
at least two $P$-wave bound states of the molecule $^7\mathrm{Li} \, 
^{87}\mathrm{Sr}_2$, whose energies should be slightly below $-5.895$ and 
$-1.138$ (in units of $^6\mathrm{Li} \, ^{87}\mathrm{Sr}$ binding energy). 
Furthermore, as  $m/m_1 > \mu_c$, one expects that a number of bound states 
and their energies depend on the details of the interactions. 

Besides influence of the three-body bound states on the properties of 
the two-component ultra-cold gases, significance of the three-body resonances 
was emphasised in~\cite{Levinsen09}. 
Furthermore, for an experimentally interesting case of 
$^6\mathrm{Li}$--$^{40}\mathrm{K}$ fermionic mixture mass ratio $m/m_1 
\approx 6.64$ is close to the value $\mu_1 \approx 8.17$ (at which 
the bound state arises), a role of the $P$-wave three-body resonance 
in scattering of $^6\mathrm{Li}$ off the $^6\mathrm{Li} \,^{40}\mathrm{K}$ 
molecule was elucidated in~\cite{Levinsen09,Levinsen11,Combescot12,Alzetto12}. 
It is worthwhile to mention also $^{173}\mathrm{Yb}$ and $^{23}\mathrm{Na}$ 
fermion-boson mixture, for which the $P$-wave resonance should be taken into 
account as $m/m_1 \approx 7.52$ is even closer to $\mu_1$. 
Possible influence of the $D$-wave resonance can be considered for 
$^{133}\mathrm{Cs}$ and $^6\mathrm{Li}$ mixture, for which 
$m/m_1 \approx 22.17$ is just below $\mu_1(2) \approx 22.34$ (corresponding to 
appearance of the $D$-wave three-body bound state). 

The effect of the $P$-wave three-body bound state in the problem of 
a light impurity atom immersed in Fermi gas was considered in~\cite{Mathy11}. 
By using the variational method, the ground-state properties were determined 
as a function of mass ratio and dimensionless scattering length and 
corresponding phase diagram was constructed. 
It was shown that for a sufficiently large mass ratio ($m/m_1 > 7$) 
the formation of a three-body molecule is energetically more preferable 
than a two-body molecule or polaron. 
As expected, on the phase diagram the boundary between regions corresponding 
to two- and three-body molecules in the low-density limit tends to the pure 
three-body result $\mu_1$. 
Furthermore, the calculated phase diagram shows that with increasing Fermi-gas 
density the formation of three-atomic molecules becomes more favourable than 
two-atomic ones. 

There is only scarce information about the properties of four two-component 
fermions. 
One of the principle results on three identical fermions and distinct particle 
was obtained in~\cite{Castin10}, where it was shown that the four-body 
$L^{P} = 1^+$ spectrum is not bounded from below for $m/m_1 \ge 13.384$. 
This means that the four-body Efimov effect takes place in the interval 
$13.384 \le m/m_1 \le 13.607$. 
Note that this peculiarity is not present for other values of total angular 
momentum and parity ($L^P = 0^+,\, 1^-$). 
Similar to the three-body 
problem~\cite{Kartavtsev07,Kartavtsev07a,Endo11,Helfrich11}, one supposes 
that the universal four-body bound states could exist below the four-body 
critical mass ratio $m/m_1 \approx 13.384$. 
Till now, there is only the calculation~\cite{Blume12}, which indicates 
existence of $L^{P} = 1^+$ bound state of three identical fermions 
and a distinct particle for $m/m_1 > 9.5$. 
Besides, one should note the interesting result on the four-body scattering 
problem~\cite{Levinsen11}, where the scattering length for two colliding 
$^6\mathrm{Li} \,^{40}\mathrm{K}$ molecules was calculated. 

Despite a marked success in describing a few two-component fermions, there are 
still many problems to be solved. 
In particular, some questions naturally arise in consideration of 
the potential-dependent states, at least in the limit $a \to \infty$. 
It is of interest to determine the smallest mass ratio, below which three 
two-component fermions are not bound by any two-body potential, and to find 
explicitly the corresponding potential. 
Likewise, it is desirable to find a maximum number of bound states, which 
could arise for mass ratio increasing up to $\mu_c$, and to determine 
the corresponding potential. 

In order to elucidate the crossover from finite to infinite number of bound 
states for $m/m_1$ around $\mu_c$, it seems to be important to take into 
account the energy dependence on the potential range $r_0$. 
In a similar way, it is of interest to trace a fate of the potential-dependent 
states (e.g., in the unitary limit $a \to \infty$) for $m/m_1$ increasing 
across the critical value $\mu_c$. 
In view of the above discussion in Section~\ref{zero-range-model} 
on application of the zero-range model in a few-fermion problem, a special 
care on the asymptotic dependence of the solution near the triple collision 
point is needed to provide a complete description in the interval 
$8.62 \le m/m_1 \le 13.607$. 

\bibliography{fermions}% Produces the bibliography via BibTeX.

\end{document}